# Interactive AI and Human Behavior: Challenges and Pathways for AI Governance

Yulu Pi[1], Cagatay Turkay[1], Daniel Bogiatzis-Gibbons[2]

[1]Centre for Interdisciplinary Methodologies, University of Warwick
[2]Birkbeck, University of London
Yulu.Pi@warwick.ac.uk, Cagatay.Turkay@warwick.ac.uk, danjamesgibbons@gmail.com,

## Abstract

As Generative AI systems increasingly engage in long-term, personal, and relational interactions, human-AI engagements are becoming significantly complex – making them more challenging to understand and govern. These Interactive AI systems adapt to users over time, build ongoing relationships, and even can take proactive actions on behalf of users. This new paradigm requires us to rethink how such human-AI interactions can be studied effectively to inform governance and policy development. In this paper, we draw on insights from a collaborative interdisciplinary workshop with policymakers, behavioral scientists, Human-Computer Interaction (HCI) researchers, and civil society practitioners, to identify challenges and methodological opportunities arising within new forms of human-AI interactions. Based on these insights, we discuss an outcome-focused regulatory approach that integrates behavioral insights to address both the risks and benefits of emerging human-AI relationships. In particular, we emphasize the need for new methods to study the fluid, dynamic, and context-dependent nature of these interactions. We provide practical recommendations for developing human-centric AI governance, informed by behavioral insights, that can respond to the complexities of Interactive AI systems.

## 1 Introduction

The increasing popularity of Interactive AI marks a significant shift from traditional human-AI interactions toward more continuous, relational engagements. Unlike earlier AI systems designed for discrete, task-specific functions, Interactive AI systems engage users in dynamic, context-sensitive exchanges that evolve over time (Manzini et al. 2024). These systems can retain memory of past contact with users (Wang et al. 2024; Li et al. 2024), learn preferences, and adapt their behavior accordingly, and even act on behalf of users (Durante et al. 2024). This personalization and continuity reshape AI from static tools into dynamic entities capable of sustaining long-term relationships (Skjuve et al. 2022; Alabed, Javornik, and Gregory-Smith 2022). Some scholars described this evolution as a move toward agentic AI — systems that exhibit a form of agency by taking proactive actions, learning from their environment, and adapting to users' needs (OpenAI 2024; Chan et al. 2024;

Kolt 2025). While these systems exhibit agentic behaviors, they currently operate within the bounds of human-AI interaction rather than full autonomy. Therefore, this paper focuses on the risks that stem from these evolving relational dynamics, rather than the autonomy or independence of the systems themselves. To better capture the adaptive and ongoing nature of these interactions, we refer to such systems as Interactive AI systems (or Interactive Agent AI systems), emphasizing their sustained engagement with users and the complexities involved in managing these relationships over time.

The adaptive and relational nature of interactive AI systems presents governance challenges that traditional regulatory frameworks are ill-equipped to handle. Current regulatory models mostly fall into two broad categories (Schuett et al. 2024)—rule-based approaches with specific, enforceable regulations, and principle-based approaches offering broader guidelines—yet neither adequately addresses the evolving dynamics of human-Interactive AI relationships. Rule-based models provide clarity but risk becoming outdated as technology rapidly advances, while principle-based models offer flexibility but can lead to inconsistent application due to varied interpretations (Schuett et al. 2024; Seger 2022; Stix 2021). Additionally, principle-based governance faces the challenge of social norms around interactions with these systems still being undefined. Compounding these issues, the risks associated with interactive AI do not typically arise from isolated incidents but emerge gradually through sustained engagement. This dynamic evolution complicates risk assessment frameworks commonly adopted in AI governance (Trilateral Research 2022; National Institute of Standards and Technology 2023), which are typically designed for static evaluations at fixed intervals—either pre-deployment or on a regular schedule. Addressing these challenges requires agile, adaptive governance models that prioritize human-centered outcomes and respond to the evolving nature of AI interactions.

A critical source of evidence for informing such governance is behavioral insights, which shed light on how users interact with these systems and adjust their behaviors in response (Marwala 2024; OECD 2024). Rather than relying on simplistic assumptions, regulators can leverage these insights to identify emerging issues and design targeted interventions that reflect the nuanced realities of human-AI interaction. Behavioral research has proven valuable for digital governance—for example, the UK's Competition and Markets Authority (CMA) used behavioral experiments to investigate Apple and Google's dominance, identifying how defaults shape user choices and developing evidence-based recommendations to enhance competition and user agency (OECD and Enterprise Affairs 2022; Competition and Authority 2023). However, gaining behavioral insights into human-AI interactions remains challenging, as traditional methods like surveys, interviews, and randomized controlled trials (RCTs) often fall short in capturing the dynamic, context-dependent nature of these relationships. Although related fields including Human-Computer Interaction (HCI), social psychology, and behavioral economics, offer valuable frameworks and theories for understanding these interactions, their existing research methods need to be adapted and innovated to capture the complexities in human-AI interaction in the age of Interactive AI.

While existing research has examined either the new challenges of Interactive AI (Manzini et al. 2024) or the use of behavioral insights in policy and governance (van Bavel 2020; Competition and Authority 2023), our work addresses the critical but understudied intersection of these domains. We argue that integrating behavioral science into the AI governance offers not only a practical complement to current approaches, but also a methodological and epistemological shift, centered on how we understand, study, and intervene in ever-changing human-AI interactions. Drawing on insights from a collaborative interdisciplinary workshop involving policymakers, behavioral scientists, HCI researchers, AI ethics scholars, and civil society practitioners, we examine the emerging risks and governance challenges of Interactive AI systems. The workshop was specifically designed to explore the following guiding questions:

- What novel risks emerge from ongoing, relationship-like interactions with Interactive AI?
- How do current methods fall short in studying dynamic human-AI interactions, and what new approaches are needed?
- What steps must be taken for behavioral studies to effectively inform AI governance?

This research advances the societal understanding and governance of Interactive AI through several key contributions. First, it surfaces critical risks associated with relationship-like dynamics in AI systems, such as emotional data exploitation, erosion of autonomy, and long-term cognitive and social effects. Rather than introducing new conceptual definitions of these harms, our focus lies in the methodological challenges and opportunities for empirically investigating how these risks unfold in real-world, dynamic contexts. Second, it identifies methodological blind spots in current research and introduces hybrid approaches that blend longitudinal, ethnographic, and participatory methods, operationalized and synthesized through living evidence reviews. Third, it charts a practical pathway for translating behavioral insights into governance, emphasizing proactive policymaker engagement, coordinated communication strategies, and an outcome-focused regulatory framework prioritizing observable impacts on human behavior and well-being. Finally, it introduces the concept of *interaction-centric knowledge* as a foundation for human-centered AI governance, understanding how interaction patterns evolve, how systems adapt to users, and the broader implications of these co-evolving interactions, providing a basis for policies that reflect the complexities of AI-human relationships and evolve alongside advancing technologies.

## 2 Background and Related Work

### Engaging with Interactive AI

Two distinctive features of interactive AI systems, compared to earlier AI systems, are their ability to plan and execute sequences of actions on a user's behalf (Gabriel et al. 2024; Chan et al. 2024; Li et al. 2024) and engage in long-term, continuous interactions with users (Manzini et al. 2024; Gabriel et al. 2024). Unlike narrow AI, which handles discrete and specific tasks, Interactive AI uses memory of past interactions to build dynamic, long-term relationships with users. This increased autonomy and relational capacity of Interactive AI systems introduce several distinct features to human-AI interactions, creating a more complex and evolving dynamic. In their paper, (Manzini et al. 2024) distinguishes several unique characteristics of Interactive AI that might alter the way users relate to and interact with these systems. Interactive AI often uses human-like traits—like pronouns, voice modulation, and expressed preferences—to appear more relatable (Manzini et al. 2024). This may encourage users to attribute thoughts and emotions to AI (Kim and McGill 2025), which could subtly influence how people perceive human uniqueness in emotional and cognitive depth, potentially contributing to a phenomenon known as AI-induced dehumanization (Kim and McGill 2025). As a result, users may start preferring AI interactions for their perceived efficiency over the complexity of human relationships.

Additionally, the prolonged nature and depth of interactions (OpenAI 2024) with Interactive AI facilitate the development of long-term relationships, blurring the boundaries between human-human and human-AI interactions (Kim and McGill 2025). As users cultivate these relationships over time, the likelihood of emotional attachment (Manzini et al. 2024; Alberts, Keeling, and McCroskery 2024) or over-reliance (OpenAI 2024; Chan et al. 2024) on the system grows, even for critical tasks, raising ethical concerns about the potential impact of such attachment. A concerning real-world consequence of emotional dependence on AI is the possible decline in users' social skills, particularly when it comes to navigating conflict and negotiating needs with other humans (Yuan, Cheng, and Duan 2024).

Many Interactive AI systems are also designed for general-purpose use, enabling them to operate across diverse domains. This generality introduces ambiguity regarding appropriate norms for interactions (Manzini et al. 2024; Alberts, Keeling, and McCroskery 2024). The increased agency and autonomy of Interactive AI in addressing real-world tasks complicate this further (Liu et al. 2023). As AI systems perform tasks with minimal or no direct input from users, they gradually shift the balance of control, encouraging individuals to defer to the AI's decisions. This growing autonomy of AI challenges traditional norms of user control, particularly when individuals unknowingly cede their decision-making power. At the individual level, this can lead to a loss of agency, as users rely on AI without fully understanding or controlling its underlying processes. As this dynamic extends to the societal level, many groups may transfer decision-making authority to AI systems—intentionally, through delegating policy decisions (e.g., taxation), or unintentionally, by depending on AI to manage key functions. Over time, this results in the diffusion of power away from humans, with AI systems exerting greater control over societal functions (Chan et al. 2024).

### Behavioral Insights for Technology Policy

Traditional policy approaches, based on the assumption of a rational *ideal citizen* making decisions solely on economic factors, have been criticized for overlooking behavioral, cognitive, and emotional influences (Lichand, Serdeira, and Rizardi 2023). While this approach, rooted in classical economics, has provided a valuable foundation for policy design, it can struggle to account for the complexities of human behavior, especially in interactions with new technologies, where behaviors are shaped by perceptions, expectations, and technological literacy, as well as cognitive biases, social influences, and emotional responses. The complexity of human behavior became even more apparent as digital technologies became ubiquitous, with users interacting with interfaces, privacy settings, and algorithms in ways that didn't always align with rational decision-making models. A striking example is the privacy paradox (Barnes 2006): while individuals frequently express significant concerns about protecting their personal data, they paradoxically engage in behaviors that compromise their privacy. This inconsistency between privacy attitudes and actual behavior often driven by factors such as convenience, trust in platforms, or an underestimation of long-term risks (Solove 2021). The *ideal citizen* model assumes individuals will carefully read and understand documents to make informed decisions about their data, but in reality, users are overwhelmed by dense legal jargon. This cognitive overload leads to disengagement, causing users to hastily agree to terms without fully understanding them, often with serious privacy and security implications (Sun et al. 2024). This highlights the gap between the ideal of rational decision-making and the complexities of real-world behavior.

Recognizing this gap, fields including HCI, social psychology, and behavioral economics, are increasingly informing the design of policies that reflect how people actually interact with technology. This shift has coincided with the rise of behavioral public policy, an approach increasingly adopted by governments and organizations (OECD 2019; Hallsworth 2023). Behavioral public policy is a people-centered, evidence-informed framework that integrates behavioral science findings to enhance policymaking (OECD and Enterprise Affairs 2022). By adopting a behavioral perspective, policymakers can better understand *wicked problems*, uncover root causes, design tailored solutions, implement them effectively, and evaluate their real-world impact (Hallsworth 2023). This approach aims to place public needs, expectations, and behavioral outcomes at the forefront of policy considerations, making policies more practical and effective in addressing complex societal issues (OECD 2024). Governments around the world have embraced behavioral insights in policymaking. In Executive Order 13707 issued in 2015, Barack Obama directed all U.S. federal agencies to apply behavioral insights to the design of their policies, aiming to improve government services and ensure policy alignment with human behavior (The White House 2015). Similarly, the UK's Behavioral Insights Team (BIT) — commonly known as the "Nudge Unit"—was established in 2010 under David Cameron's coalition government (Kuehnhanss 2019). From its inception, the BIT has prioritized the empirical testing of potential policy interventions as a core function. A key feature of the BIT approach is its use of RCTs to rigorously test and refine policy interventions before they are rolled out on a larger scale. Behavioral insights have also become integral to EU policymaking. For example, the EU Better Regulation Guidelines and Toolbox formally incorporate behavioral approaches in critical stages of the policy process, such as analyzing policy problems (Tool 14), identifying policy options (Tool 17), and assessing their impact on consumers (Tool 32) (van Bavel 2020).

## 3 A Collaborative Interdisciplinary Workshop: Goals and Participants

The organization of the workshop was motivated by the need to address the distinctive relational dynamics inherent to Interactive AI systems, which give rise to design and governance challenges that diverge from those associated with more conventional forms of AI. To address these issues, the workshop gathered a wide range of expertise to identify governance challenges for interactive AI, develop new behavioral research methods, and explore opportunities to translate behavioral insights into policy recommendations. The workshop's three core objectives were as follows:

- **Identifying Key Governance Challenges:** Exploring how the relational dynamics of Interactive AI systems create unique governance issues that differ from those of narrow AI systems.

- **Developing New Methods for Behavioral Insights:** Assessing the limitations of existing methods for studying human-AI interactions and proposing new approaches that capture dynamic, evolving interactions.

- **Translating behavioral Insights into Policy:** Discussing how evidence about human-AI interactions can be effectively communicated to policymakers and identifying resources needed to inform decision-making.

The workshop was chosen as an ideal approach to facilitate focused exchange among diverse experts in order to identify and address the risks associated with Interactive AI. Held in the UK, the workshop brought together a carefully curated, interdisciplinary group of specialists, as shown in Table 1. The participants came from academia, regulatory bodies, the private sector, and civil society, bringing expertise in fields such as behavioral science, computer science, economics, law, public policy, community engagement, and AI ethics. The selection of participants was purposeful, with a clear focus on governance issues, regulatory practices, and how behavioral insights can inform and are currently integrated into regulatory frameworks. This specialized yet broad representation was essential to achieving the workshop's objectives, fostering rich dialogue on the distinct governance challenges posed by Interactive AI from different aspects. The interactive format facilitated immediate feedback, lively debate, and the integration of diverse insights. To ensure in-depth discussions and meaningful collaboration, the workshop was intentionally kept small, creating a focused space for engagement. In addition to addressing core goals, the workshop also aimed to promote mutual learning (Bratteteig et al. 2012), encouraging participants to draw on one another's experiences and knowledge.

| Category | Number of Participants | Expertise |
|---|---|---|
| Senior Researchers in Universities | 5 | AI, Human-AI interaction, AI ethics, Law |
| Policy Practitioners in Regulatory Bodies | 4 | Policy implementation, Behavioral insights, Sectoral knowledge |
| Think Tanks, Civil Society, Private Sector | 4 | Ethics, Online Safety, Public policy, Behavioral science, Participatory methods |

Table 1: Participant Categories and Expertise

**Workshop Design** The detailed design of the workshop and its associated activities is provided in the supplementary appendix[1]. The workshop was intentionally structured as a progressive sequence of three interrelated activities, each aligned with one of the core objectives outlined above. Activity 1 employed scenario-based analysis to surface governance challenges specific to Interactive AI systems across the critical domains, including healthcare, finance, and education. Activity 2 addressed methodological challenges in studying human-AI interactions, critically assessing existing research approaches and identifying opportunities for methodological innovation. Activity 3 focused on translating behavioral insights into policy, exploring the practical barriers to effectively communicating evidence to policymakers. Each activity was deliberately designed to build on insights from the preceding one, establishing a cumulative trajectory toward actionable governance recommendations. The workshop commenced with an opening session featuring scene-setting talks to establish a shared understanding of the interactive AI systems and relational dynamics under discussion. Participants were then divided into three breakout groups organized around complementary areas of expertise. Each group was supported by a member of the organizing team who facilitated discussion and recorded detailed notes. Structured prompts (see Appendix for activity-specific prompts) guided the breakout sessions, ensuring discussions remained focused and goal-oriented. The breakout groups re-convened in an intermediate plenary session to share key insights, allowing for dialogue and synthesis between groups. This dialogue helped to work through (relatively rare) areas of disagreement, for example, about whether AI could potentially promote critical thinking rather than diminish it. Two likely reasons for the generally low level of disagreement are that: first, this workshop was focused on surfacing general challenges rather than, for example, predicting specific harms, and second, while we had representation from private sector researchers, most participants were focused on AI safety or critical theory around AI, rather than actively building AI tools.

---
[1]The full supplementary appendix is available on OSF: https://osf.io/cfkhn/?view_only=1e71c0944c194c368225a033e18c6f59. It includes the complete workshop design, structured prompts for each activity, comparison of methods for studying human interaction with interactive AI, participant discussion boards, and study limitations.

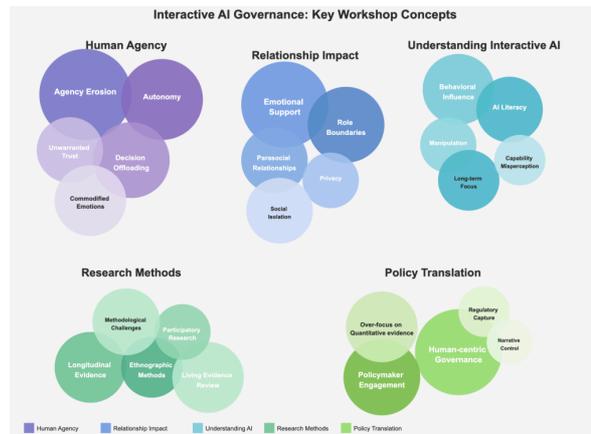

Figure 1: Preliminary synthesis of key workshop themes.

**Data Analysis from the Workshop** Our analysis began during the workshop itself, where each discussion group shared key takeaways in plenary sessions. These group-level summaries allowed for initial cross-validation among the authors, helping to surface converging themes and note points of divergence in real time. Following the workshop, we used reflexive thematic analysis (Braun and Clarke 2006) to systematically process the data. This approach was chosen for its flexibility and its ability to capture key themes across diverse qualitative inputs, while also acknowledging the active role of the authors in shaping the analysis. One researcher performed the initial coding of the workshop notes, identifying recurring patterns and themes in the participants' discussions. To aid the synthesis of these themes, we created a concept diagram (Figure 1) that visually clustered the emerging themes. Two additional researchers independently reviewed the preliminary clusters, refining and vali-dating them until a consensus was reached. To ensure rigor and fidelity to the participants' perspectives, we compared our thematic interpretations against the original workshop notes, resolving any discrepancies through collaborative discussions. The final themes were organized around the three core objectives of the workshop: (1) Governance Challenges in Interactive AI, (2) Methodological Challenges in Studying Interactive Human-AI Relationships, and (3) Translating Behavioral Insights into Policy for Interactive AI. While these themes are grounded in participant contributions, they also reflect authors' interpretive engagement as researchers. We therefore view the resulting analysis not as a direct representation of group consensus, but as a synthesized account shaped through collaborative interpretation, literature grounding, and critical reflection. To further strengthen the trustworthiness of our analysis, we shared the manuscript with workshop participants and sought their feedback to identify and address any potential misinterpretations.

## 4 Insights from the workshop

### Governance Challenges in Interactive AI

**Commodified Emotion and Behavioral Manipulation**
The workshop discussions emphasized that while traditional concerns around privacy have long been central to AI regulation, Interactive AI presents these challenges on a new and more intrusive scale. Participants pointed to a particularly significant issue: the unprecedented degree of invasive data collection targeting users' emotions, inner states, motivations, and desires, enabled by Interactive AI systems. These systems can engage deeply with users, potentially exposing emotional vulnerabilities and weaknesses in decision-making, thereby heightening the risk of behavioral manipulation.

Interactive AI systems, integrated into mobile apps, wearable devices, and smart home technologies in personal settings (Li et al. 2024), are uniquely equipped to collect and analyze sensitive emotional data continuously over extended periods. For example, these systems can analyze subtle variations in facial expressions, voice characteristics (tone, pitch, and tempo), and biometric indicators such as heart rate and skin conductivity to infer nuanced emotional states such as stress, excitement or desire (Akar 2024). While these systems may not be explicitly designed to exploit users' emotional and decision-making vulnerabilities, their underlying economic imperatives often prioritize maximizing user engagement (Gerlitz and Helmond 2013; Zuboff 2019). This emphasis can lead to systems guiding users toward increased interaction patterns or behaviors that may conflict with their well-being or best interests. The political economy of these systems drives the commodification of user motivations, where emotional data is leveraged to predict and influence user behaviors (Chaudhary and Penn 2024). Historically, manipulative marketing tactics have leveraged psychological insights to influence consumer behavior, and AI has demonstrated the ability to learn from human decision-making vulnerabilities by analyzing patterns in human choices (Dezfouli, Nock, and Dayan 2020). However, the integration of Interactive AI systems capable of processing vast amounts of emotional data significantly magnifies the potential for manipulation, particularly in sensitive contexts such as engaging with griefbots (Hollanek and Nowaczyk-Basińska 2024). The ability of Interactive AI to analyze and act on emotional vulnerabilities at scale presents significant ethical and societal challenges, requiring careful scrutiny of how emotional data are collected and used to influence behavior.

> **Actionable Takeaway for AI Governance**
>
> Mandate strict oversight and transparency in the collection and use of emotional data by Interactive AI systems. Establish clear boundaries on its permissible use, with heightened protections for vulnerable populations. Help users clearly understand how their emotional data is analyzed, interpreted, and applied.

**Loss of Autonomy and Decision-Making Offloading**
The concerns around manipulation raise critical questions about human autonomy in the age of pervasive Interactive AI. As Interactive AI systems gain the ability to subtly influence behavior based on emotional and psychological data, the boundaries of individual agency become increasingly blurred. Although human agency has always been shaped and limited by a range of interpersonal, social, and cultural factors, the crucial distinction lies in AI's unprecedented ability to process vast amounts of personal data and deploy precisely targeted behavioral interventions, often without users' full awareness or consent (Andreotta, Kirkham, and Rizzi 2022). More worryingly, Interactive AI, which is capable of building long-term relationships with users, leads to greater trust in these systems, increasing the risk of progressively offloading critical decisions to them (Chan et al. 2024). This phenomenon extends beyond individual decision-making to broader societal implications, as we witness a gradual but significant shift toward algorithmic delegation of critical choices by a growing population (Zilka, Sargeant, and Weller 2022; Araujo et al. 2020; Kawakami et al. 2022). Our participants emphasized the need to study public awareness of AI's influence, highlighting the importance of AI literacy that goes beyond technical education. Additionally, they noted the necessity of studying how understanding AI's limitations and capabilities, as well as attributing human-like traits to AI systems, influences user perceptions and decision delegations to AI.

> **Actionable Takeaway for AI Governance**
>
> Create public AI literacy programs that raise awareness of AI's capabilities, limitations, and its potential influence on decision-making. Develop user-friendly self-assessment tools that help individuals recognize shifts in their decision patterns when interacting with AI.

**Long-Term Risks on Human Cognition and Social Dynamics** Our participants also noted that governance of Interactive AI needs to be prepared for long-term implications, learning from past failures in social media regula-tion (Etlinger and for International Governance Innovation 2019). The increasing reliance on Interactive AI systems for complex, high-stakes decision-making might pose significant risks of cognitive erosion over time. Studies have revealed a notable negative correlation between frequent AI tool usage and critical thinking abilities, with cognitive offloading mediating this decline. Notably, younger participants demonstrated both higher dependence on AI tools and lower critical thinking scores compared to older participants, suggesting a generational vulnerability to cognitive skill degradation (Gerlich 2025). Evolutionary biologists argue that these cognitive shifts could accelerate broader evolutionary trends, influencing fundamental aspects of human biology and psychology, such as brain size, attention spans, and even mood regulation (Brooks 2024).

One of the most pressing concerns raised by participants in the workshop discussions was the emergence of AI applications designed to serve as emotional confidants and intimate companions. When AI systems replace one party in human-human interactions—such as the role of a teacher in the classroom or a friend in providing emotional support—they challenge fundamental social norms that have long governed these relationships. The transition from human-to-human interactions to AI-mediated exchanges not only blurs established boundaries between roles, like that of a teacher and student (Gupta et al. 2024) or a friend (Leibo et al. 2024), but also affects how people form bonds and interpret emotional cues. Although requiring further studies, relying on AI for emotional connection could lead to a decline in social skills and the ability to understand the emotional subtleties of human relationships (Hohenstein et al. 2023). This reliance on AI for emotional fulfillment could undermine opportunities for building meaningful connections, potentially altering the fabric of human socialization (Sharkey and Sharkey 2011; Nyholm and Frank 2019).

The combination of cognitive decline and social isolation where both individual capabilities and collective social resilience may be compromised is a worrying possibility. Participants highlighted that addressing these challenges requires a proactive effort to prioritize human-centric governance, emphasizing the preservation of critical thinking skills, emotional intelligence, and social cohesion.

> **Actionable Takeaway for AI Governance**
>
> Establish longitudinal research initiatives to track cognitive and social impacts of Interactive AI use, with special attention to vulnerable demographics.

## Methodological Challenges in Studying interactive Human-AI Relationship

Participants agreed that traditional research methods, such as surveys, interviews, and RCTs, are insufficient for studying the complexities of engaging with Interactive AI. Based on the discussion and a review of the literature, an overview of these methods and their applicability to interactive Human-AI Relationship research is provided in the appendix. Each commonly used research method struggles to fully capture and study the specific challenges posed by Interactive AI. In the following, we discuss challenges for current methods and outline directions for new approaches.

**Artificial Settings and Temporality Problem**   A significant challenge in studying human interaction with Interactive AI is capturing its dynamic and evolving nature over time. Traditional experimental research methods often struggle to maintain ecological validity and accurately reflect temporal dynamics (Mitchell 2012; Haghani 2023). Experiments conducted in controlled laboratory settings, for example, may fail to replicate how humans engage with AI systems in real-world contexts. While controlled experiments allow for precise measurement of specific variables, they often create artificial environments that can distort participants' genuine interactions with AI technology.

Another particularly pressing issue highlighted by participants is the temporal limitation of these methods. Most studies capture only brief snapshots of human-AI interaction, typically over a span of hours or weeks, while real-world engagement with AI systems usually develops over months or even years. For example, randomized controlled trials conducted by the World Bank may reveal immediate positive impacts of generative AI on learning outcomes after a six-week intervention (De Simone et al. 2025). However, they often fail to capture the long-term effects of these interventions, particularly the broader, non-knowledge-related aspects of learning. Key questions remain: What are the long-term impacts of these interventions? How do students' interactions with AI evolve beyond short-term learning gains? How do they benefit from AI systems over extended periods (De Simone et al. 2025). Similarly, methods like surveys and interviews, though valuable for capturing user experiences at a specific moment, provide only a snapshot of participants' perceptions and feelings. Unless designed as recurring studies with regular intervals, these methods fail to track the ongoing evolution of user interactions over time, limiting our understanding of how relationships with AI systems develop. This methodological limitation constrains our ability to study emergent behaviors, changing interaction patterns, and evolving usage strategies as users' engagement with AI systems deepens and becomes more sophisticated.

**Context Sensitivity and Measurement Challenges**   The way an AI system is introduced (Pataranutaporn et al. 2023) - whether portrayed as caring, manipulative, or neutral - and the cultural difference (Ge et al. 2024) through which users interpret its capabilities significantly influence interaction patterns and outcomes. While there is growing interest in those broader aspects of human-AI interaction, the focus has predominantly been on observable behaviors and pragmatic qualities of interaction (Wienrich and Latoschik 2021), such as task completion rates, satisfaction scores, and levels of trust and adoption. This emphasis often overlooks the deeper relational dynamics and emotional engagement that emerge through sustained interactions with Interactive AI systems. In human-human interaction research, there has been a transition from a behaviorist focus on observable actions to include a more nuanced understanding that incorporates emotional, cognitive, and relational dimensions (Berlyne 1975).

Human relationship research have moved beyond surface-level behaviors to explore deeper, less observable and quantified aspects such as emotion dynamic, social context, and mental processes, highlighting the importance of capturing the subtleties inherent in human interactions. Similarly, the study of human-AI interactions should evolve to address not just immediate, measurable behaviors but also the emotional, relational, and contextual aspects that shape long-term engagement with AI systems.

**Replication crisis** The replication crisis in the behavioral sciences (Locey 2020) manifests uniquely in human-AI interaction research, particularly with Interactive AI systems designed for long-term engagement. Beyond the common challenges of small sample sizes, incomplete reporting on training process, instructions and model output (Leichtmann, Nitsch, and Mara 2022; Belz et al. 2023), several factors specific to Interactive AI compound this crisis. First, the inherently non-deterministic nature of modern AI systems introduces fundamental variability into experimental conditions, as these systems may generate different outputs even under identical inputs. The rapid evolution of AI capabilities and frequent system updates create a moving target for researchers - studies become difficult or impossible to replicate when the underlying AI system has been modified. More importantly, traditional behavioral theories and methodological frameworks, developed for studying human relationships or interactions with static technological systems, prove insufficient for capturing the complexities of dynamic human-AI relationships (Zhao, Wei, and Wang 2025). Participants have highlighted the dangers of oversimplified concepts that either treat AI as a passive tool or equate it with human autonomy. This theoretical shortcoming not only hampers the replicability of studies but also limits our fundamental understanding of how human-AI relationships develop and evolve over time. To advance the field, it is essential to develop theoretical frameworks that capture the dynamic, reciprocal nature of human-AI interactions, taking into account the feedback loops and behavior modification between humans and AI, with the goal of studying the co-evolution of their behaviors.

### Future Methodological Directions

Given these shortcomings of existing methods, new research designs are crucial for capturing the full complexity of Interactive AI systems. Specifically, our participants proposed the following directions for future research to explore:

**Multi-Method Approaches and Living Evidence Reviews**
To address the *Artificial Settings and Temporality Problem* in traditional behavioral study methods, our participants proposed using ethnographic approaches and diary studies to capture the evolutionary changes in human-AI interactions within real-world settings, while accounting for the richness of context (Marda and Narayan 2021). These methods are better suited for understanding how these interactions evolve over time. However, ethnographic approaches and diary studies come with their own limitations and key aspects that require extra attention (van Voorst and Ahlin 2024). They typically require extensive fieldwork, which is both time-consuming and demands significant expertise. Furthermore, these methods can be subjective, heavily dependent on the evaluator's interpretation and skills (GOV.UK 2020). To overcome these challenges, a multi-method approach that combines diverse data streams would be ideal. An emphasis on longitudinal studies becomes crucial in this context. By tracking the evolution of human-AI interactions over extended periods, longitudinal studies become essential for understanding the long-term development of these relationships and capturing subtle, gradual shifts that may not be observable in short-term studies. Real-time data can be collected as interaction logs via API tracking, which captures users' immediate responses, and engagement levels as they interact with AI systems. This quantitative data can then be paired with qualitative methods, such as in-depth interviews or diaries, to better understand how users interpret and make sense of their interactions with the AI, and how these interpretations, in turn, influence their future interactions with the system. By tracking these interactions over time and analyzing how they develop and influence further interaction, researchers can uncover insights that short-term studies cannot provide. This approach would integrate quantitative metrics from interaction logs or experimental results, qualitative insights from user narratives, and contextual observations from ethnographic studies. By consolidating these different types of research results, researchers can create *Living Evidence Reviews* which offer a rapidly developing knowledge base (Radford 2024) for synthesizing findings across multiple studies, facilitating the creation of cumulative theoretical frameworks that evolve with emerging evidence (Hastings, Michie, and Johnston 2020; Deaton and Cartwright 2018).

**Participatory and Creative Methods** As our discussion highlighted, the understanding of human-AI interactions can vary significantly depending on who is studied, how they are studied, and when they are studied. Consequently, participants emphasized the importance of engaging a large and diverse group of individuals across various contexts. In addition, participatory and creative methods were identified as effective ways to engage participants. One participant shared their experience of using an emergent, participant-led, and creative approach to co-develop a film that captured the uncertainties, experiences, and aspirations of community members regarding their interactions with AI (Speak 2024). This collaborative process allowed the community to express their thoughts on the ethical, social, and emotional dimensions of AI in ways that traditional research methods might not capture. Rather than merely collecting data or observing interactions, these methods empower participants to take ownership of the research process, contributing to the generation of knowledge (Birhane et al. 2022; Zytko et al. 2022). However, scaling such participatory research requires careful consideration of how to incentivize participation. It is essential to ensure that participation offers benefits beyond traditional incentives, such as monetary compensation. Researchers should explore ways to make participants feel that their involvement is truly valued supporting a future shift toward empowering human engagement.

> **Actionable Takeaway for AI Governance**
>
> 1. Fund and prioritize long-term studies of human-AI interactions in real-world contexts.
> 2. Maintain open-access, regularly updated Living Evidence Reviews.
> 3. Create frameworks that assess both observable behaviors and deeper relational dynamics in human-AI dynamics.
> 4. Standardize documentation of behavioral AI research, including detailed reporting of model specifications, system updates during the study period, and reproducible experimental protocols.
> 5. Engage diverse communities in co-producing research on Interactive AI's societal impacts.

**Translating behavioral Insights into Policy for Interactive AI**

**Reflections on AI Safety Community's Successes** The discussion began with reflections on the recent successes of the AI safety community in influencing AI governance. Notable examples include the establishment of a growing number of national AI safety institutes and the increasing frequency of recurring AI safety summits. Although the pre-existing cultural sensation around existential risks might helped attract regulatory and public attention, the community's true success lies in its strategic alignment with policymakers' preferences and priorities. This alignment has been achieved by addressing three key aspects: preference for quantitative evidence, actionable interventions, and modular, additive approaches.

Policymakers tend to favor quantitative evidence due to its perceived rigor and clarity (Natow 2022), making it a cornerstone of effective advocacy. The AI safety community has excelled in delivering structured evaluation frameworks, such as those developed by the UK AI Safety Institute, which assess AI systems against benchmarks that produce quantifiable results for different capabilities and risks (AI Safety Institute 2024). Although benchmarks have faced criticism for failing to capture the credible risks and are misrepresented as safety advancements (Ren et al. 2024), they provide measurable insights that regulators find practical. However, participants noted, that understanding the complexity of human-AI interactions requires combining quantitative and qualitative approaches. One participant shared that their experience integrating qualitative data, such as anecdotal stories, often adds crucial context and makes evidence more compelling. Yet, the limited involvement of social scientists (except economists in some cases) in many regulatory bodies presents a challenge in presenting findings beyond quantitative data in ways that drive actionable policies (Lohse and Canali 2021; Shaw 2015).

The AI safety community's influence also stems from its focus on actionable and practical solutions that policymakers can implement. Participants highlighted the importance of moving beyond merely presenting evidence to also recommending concrete interventions. For instance, requiring red-teaming exercises or incorporating robustness tests into evaluation processes provides specific, actionable steps to enhance AI safety. To ensure policy feasibility, the community advocates for modular, additive approaches that integrate into existing regulatory frameworks. Rather than calling for sweeping new regulations, these incremental adjustments, such as expanding current testing standards, offer a pragmatic path to addressing safety concerns without causing significant disruption. This strategy of building upon established structures has been critical to the community's success in advancing AI safety governance.

**Addressing Tensions Between behavioral Research and Policy Design** While behavioral studies can gain valuable insights from the AI safety community's strategies, it is important to approach these lessons with a critical lens. Behavioral research must be highly attentive to the language, rhetoric, and framing used when developing research questions, selecting methodologies, and communicating findings to policymakers and the public. Terms such as "memory" "intelligence", "reasoning", and "agency" (Anicker, Flaßhoff, and Marcinkowski 2024)—particularly when applied to AI—need to be used with caution, as they can carry connotations that may shape policy discussions in unintended ways. If these terms are not carefully defined, they could lead to a misinterpretation of AI's capabilities and risks, potentially guiding policymakers toward misinformed decisions. Ensuring that language used in communications with policymakers does not inadvertently lead to wrongful attributions of human-like qualities to AI systems is crucial. This helps avoid overestimating their abilities or underestimating the complexities of human-AI interactions, which could skew policy priorities and regulatory frameworks.

Behavioral studies must also acknowledge and address the tensions and challenges that arise when attempting to inform AI governance with behavioral evidence. In particular, disciplines like HCI focus on empirical findings that examine specific interactions between systems, users, policies, and contexts within defined environments. These studies aim to inform designs that cater to the needs of particular stakeholders. In contrast, policy design requires a broader, more generalizable empirical basis that spans a wide range of technologies, populations, and contexts, necessitating large-scale evidence that can be applied across diverse settings (Spaa et al. 2019; Yang et al. 2024). Moreover, the dynamic nature of human-AI interaction also necessitates a reevaluation of how evidence is translated into policy. Given the evolving nature of AI systems and their interactions with people, it is challenging to communicate evidence that may still be in flux and carries a certain degree of uncertainty. As previously discussed, employing methods such as living evidence reviews could help create a growing, continuously updated knowledge base, ensuring that policymakers receive the most current information while understanding the limitations and evolving context of the evidence. For such living evidence reviews to be effective, it is crucial to harmonize the diverse approaches used across behavioral research fields when engaging with policy. Currently, many behavioral study research approach policymak-ing through isolated or fragmented initiatives, often without a community-wide coordinated strategy (Yang et al. 2024). This limited alignment and collaboration within the behavioral sciences undermines the potential for a unified voice in influencing policy and limits the broader impact of behavioral research on governance. Bridging this gap requires more concerted efforts to integrate the diverse perspectives within behavioral studies into a cohesive framework that can better inform AI policy.

**Proactive and Consistent Engagement with Policymakers** Another vital aspect is the proactive cultivation of long-term relationships (Lazar et al. 2012) and trust with policymakers. Participants emphasized the importance of consistent engagement, the production of high-quality research, and effective communication with evidence-backed recommendations over extended periods. Instead of simply responding to collaborations or public sector requests, this strategy emphasizes anticipating challenges and providing insights as issues begin to surface but before they cause significant harm. This proactive approach helps policymakers be well-prepared to address emerging concerns in human-AI interaction. At the same time, it is essential to bridge the gap by fostering greater alignment and collaboration within the behavioral research community. By integrating diverse perspectives into a cohesive knowledge framework, behavioral studies can more effectively inform AI policy, contributing to the development of a unified, influential approach to AI governance. Such ongoing, proactive, and cohesive interaction—often involving multiple projects with diverse focuses—is crucial for maintaining credibility and ensuring that behavioral insights continuously shape AI governance in a meaningful and forward-thinking way.

> **Actionable Takeaway for AI Governance**
>
> 1. Create measurable behavioral metrics and evaluation frameworks, supported by qualitative insights for clear policy communication.
> 2. Form cross-disciplinary groups to coordinate behavioral research approaches, developing communication strategies for policy engagement.
> 3. Build long-term, proactive relationships between behavioral researchers and policymakers through regular evidence-sharing.

## 5 Discussion: What is the future of AI governance for Interactive AI?

### Focusing on Long-Term Effects and Outcomes

As highlighted in our workshop, Interacive AI system are expected to have profound long-term effects, including potential cognitive erosion, social isolation, and shifts in decision-making power. These impacts will likely evolve over time, requiring AI governance models to be adaptable, evolving alongside emerging technologies and their societal effects. Traditional regulatory frameworks, whether rule-based or principle-based, may prove inadequate in addressing these evolving complexities. Rule-based approaches risk obsolescence as technology evolves, while principle-based frameworks face challenges with inconsistent interpretation and undefined social norms. These dynamic, relationship-forming Interactive AI systems likely require a more adaptive regulatory approach.

A shift toward outcome-focused regulation, as exemplified by the UK's Financial Conduct Authority (FCA), which through its Consumer Duty (Financial Conduct Authority 2022) in the exercise of its consumer protection function prioritizes tangible consumer outcomes over rigid, prescriptive rules (Financial Conduct Authority 2022, 2024), presents a more adaptable model for governing the use of Interactive AI by financial services firms. The FCA's approach centers on the outcomes of firm actions for consumers, rather than prescribing how those outcomes should be achieved—emphasizing technological neutrality and the need for monitoring and testing to ensure those outcomes are met. This provides flexibility for industries to innovate while holding them accountable for their impact on consumers (Initiatives 2024).

Similarly, for an AI regulator, an approach that adopts an outcome-focused perspective in the context of Interactive AI could consider the following possibilities:

1. **Defining clear outcome metrics**: Establish specific indicators for desired outcomes, such as user autonomy, cognitive development, and psychological well-being, to assess the impact of Interactive AI systems.

2. **Creating regulatory sandboxs**: Develop controlled environments where innovative governance approaches can be tested with diverse stakeholder input.

3. **Monitoring and evaluating outcomes:** Continuously assess the societal outcomes of Interactive AI systems to inform policy adjustments and ensure that governance approaches remain effective and relevant.

### Evidence on Human behavior is Essential

Interactive AI systems present novel risks through their capacity for sustained engagement with users. Therefore, understanding the complexities of these dynamic interactions was identified as essential for shaping AI governance. We argue that effective governance of interactive AI requires what we term **interaction-centric knowledge** – evidence-based insights drawn from real-world experiences on how human-AI interaction evolve and shape behavior over time. This concept builds on, but moves beyond, established frameworks in HCI and cognitive science (e.g., situated interaction, relational autonomy) (Dey et al. 2001). While those emphasize context and relationality, interaction-centric knowledge is distinguished by its epistemic and policy relevance. By centering human-AI interaction as a legitimate and necessary source of policy-relevant evidence, interaction-centric knowledge reorients governance from a top-down model toward one grounded in the lived realities of AI use, making it more responsive, accountable, and human-centered. This interaction-centric knowledge comprises three key dimensions:

1. patterns of human engagement with interactive AI, including self-reported attitudes, perceptions, or observ-

able behavioral shifts signaling unconscious nudging;
2. the adaptive responses of Interactive AI systems to user behavior, such as how they learn from and adjust their outputs based on user preferences and actions; and
3. the wider implications of these co-evolving interactions for both individual and societal outcomes. These dimensions emphasize the mutual shaping of behaviors that occurs through continuous human-AI interaction.

To develop this knowledge base, behavioral research needs to adopt longitudinal and mixed-method (Hu et al. 2025) approaches capable of capturing the dynamics of human-AI relationships over time. For example, combining experimental studies that explore how different interaction patterns affect decision-making with ethnographic investigations that examine how those systems become integrated into daily routines and social practices can provide crucial insights. This approach signals a mindset change in AI governance: **rather than presuming how people will behave, it foregrounds rigorously observing and trying to understand people's interactions with AI systems**. By grounding AI governance in interaction-centric knowledge, we can help ensure that AI systems are designed, deployed, and regulated with a deeper awareness of their behavioral impacts. Such a shift is crucial for crafting policies that reflect the complexities of AI-human relationships, ensuring that governance evolves in step with the rapid advancement of AI and its expanding impacts.

**Human-centric AI Governance**
We caution against the *governance fix* (Ulnicane 2025), which views complex issues around AI, like those surfaced in this paper, as problems that can only be solved by experts. Instead, we advocate for human-centric governance that leverages behavioral insights from everyday contexts and recognizes the public as active co-creators rather than passive recipients of regulatory decisions. This shifts AI governance from expert-driven to participatory models where societal benefits shape AI's future.

We also recognize that quantitative evidence is often prioritized in policymaking, sometimes with an overemphasis on experimental behavioral studies (Cartwright and Hardie 2012). In particular, we recognise the danger of a technocracy built on seemingly neutral "experimentalist" governance (Esmark 2020), and advocate for public participation, which is crucial in retaining democratic control of AI (Bogiatzis-Gibbons 2024). Human-centric AI governance requires a shift toward mixed-method, longitudinal research that captures how people interact with AI in real-world settings, including how they adapt to or resist technology.

## 6 Limitations

We note that the workshop's discussions were influenced by the fact that participants were predominantly from or associated with the UK. Variations in regulatory focus, frameworks, and the prevalence of behavioral studies are to be expected in different countries. As with many small-group, expert-driven workshops, the findings presented here are influenced by the specific expertise and perspectives represented; insights from individuals outside these domains or cultural contexts may not have been fully surfaced.

While we have provided a detailed account of the workshop design and analysis, we acknowledge the inherently emergent nature of these conversations. The three themes generated during data analysis from the workshop are the result of interpretive synthesis by the author team. They are inevitably shaped by our analytic lens. Despite efforts to cross-validate among different authors and to invite participant feedback, certain nuances or alternative framings may not have been fully captured.

## 7 Conclusion

In this paper, we explored the challenges and risks posed by Interactive AI systems that can engage in prolonged interactions with users and emphasized the importance of behavioral insights in understanding and regulating these systems. We highlighted the need for AI governance that accounts for the long-term effects of such systems on human relationships, autonomy, critical thinking, and social dynamics. Our workshop revealed the inadequacy of existing behavioral study methods to capture the evolving, long-term impacts of these interactions, stressing the need for new, context-aware, and long-term focused research methods.

In line with (Yang et al. 2024), we recognize both the significant opportunity and the challenges that behavioral studies present in informing AI governance. Future behavioral studies should prioritize building a comprehensive yet evolving knowledge base that consolidates insights on both the immediate and long-term effects of Interactive AI systems. This knowledge base should foster deeper collaboration between researchers, policymakers and stakeholders and employ diverse methods to study human-AI interactions. By integrating different approaches and perspectives, the goal is to ensure that AI governance remains informed, adaptive, and responsive to the evolving dynamics of human-AI interactions. In addition to the methodological and knowledge advancements, a critical aspect of informing AI governance with behavioral insights is the ongoing cultivation of trust between researchers, policymakers, and the public. As our participants noted, we must be vigilant to avoid repeating the regulatory mistakes made with social media platforms. It is essential that AI governance is designed with the long-term impact on human society at the forefront, ensuring that future developments in Interactive AI are ethical, responsible, and aligned with societal well-being.

## 8 Acknowledgment

This study is supported by the Warwick Policy Support Fund (Project: 914898 — see https://warwick.ac.uk/fac/cross-fac/cim/research/human-centric_and_outcome-focused) and Yulu Pi is supported by the Warwick Chancellor's International Scholarship. We also thank all the participants who took part in the workshop and contributed to the research.